\documentclass[]{aa} 
\usepackage{graphicx} 
\usepackage{natbib,color} 
\bibpunct{(}{)}{;}{a}{}{,} % to follow A&A style 
\newcommand{\HII}{\ion{H}{ii}} 
\newcommand{\HI}{\ion{H}{i}}

\newcommand{\lsun}{L$_{\odot}$} 
\newcommand{\msun}{M$_{\odot}$}

\newcommand{\degree}{\ensuremath{^\circ}}
\begin{document}  

\title{A complete $^{12}$CO 2--1 map of M51 with HERA: \\
  II. Total gas surface densities and gravitational stability}
    
   \author{  
     M.\,Hitschfeld\inst{1} \and  
     C.\,Kramer\inst{1,2} \and  
     K.F.\,Schuster\inst{3} \and  
     S.\,Garcia-Burillo\inst{4} \and  
     J.\,Stutzki\inst{1} 
          }   
     
   \institute{  
     KOSMA, I. Physikalisches Institut, Universit\"at zu K\"oln,  
     Z\"ulpicher Stra\ss{}e 77, 50937 K\"oln, Germany   
     \and  
     IRAM, Avenida Divina Pastora 7, N$\acute{\rm u}$cleo Central, E-18012
     Granada, Spain   
     \and
     IRAM, 300 Rue de la Piscine, F-38406 S$^t$ Martin d'H\`{e}res, France  
     \and  
     Observatorio de Madrid, Alfonso XII, 3, 28014 Madrid, Spain
    }  
  
   \offprints{M.\,Hitschfeld, \email{hitschfeld@ph1.uni-koeln.de}}  
   \date{Received / Accepted }  
     
% context, aims, deed, results, conclusions 
   \abstract
   % Context.
   { To date the onset of large-scale star formation in galaxies and
     its link to gravitational stability of the galactic disk have not been
     fully understood.  The nearby face-on spiral galaxy M51 is an
     ideal target for studying this subject.}  
   % Aims.
   { This paper combines CO, dust, \HI, and stellar maps of M51 and
     its companion galaxy to study the H$_2$/\HI\ transition, the
     gas-to-dust ratios, and the stability of the disk against
     gravitational collapse. }
%   { This paper discusses the parameters for gravitational stability
%     following the Toomre analysis. The total gas density
%     distribution, the velocity dispersions of the molecular, atomic
%     and stellar components are presented.}  
   % Methods.
   { We combine maps of the molecular gas using $^{12}$CO 2--1 map
     HERA/IRAM-30m data and \HI\ VLA data to study the total gas surface
     density and the phase transition of atomic to molecular gas. The total
     gas surface density is compared to the dust surface density
     from 850\,$\mu$m SCUBA data. Taking into account the velocity
     dispersions of the molecular and atomic gas, and the stellar
     surface densities derived from the 2MASS $K$-band survey, we derive
     the total Toomre Q parameter of the disk.  }
   % Results.
   { The gas surface density $\Sigma_{\rm gas}$ in the spiral arms is
     $\sim 2-3$ higher compared to that of the interarm regions. The
     ratio of molecular to atomic surface density shows a nearly
     power-law dependence on the hydrostatic pressure $P_{\rm hydro}$.
     The $\Sigma_{\rm gas}$ distribution in M51 shows an underlying
     exponential distribution with a scale length of h$_{\rm gas}=7.6$
     kpc representing 55\% of the total gas mass, comparable to the
     properties of the exponential dust disk. In contrast to the velocity
     widths observed in \HI, the CO velocity dispersion shows enhanced
     line widths in the spiral arms compared to the interarm regions.
     The contribution of the stellar component in the Toomre
     Q-parameter analysis is significant and lowers the combined
     Q-parameter Q$_{\rm tot}$ by up to 70\% towards the threshold for
     gravitational instability. The value of Q$_{\rm tot}$ varies from 1.5-3 in
     radial averages. A map of Q$_{\rm tot}$ shows values around 1
     on the spiral arms indicating self-regulation at play. }
   % Conclusions.
     {} 
     \authorrunning{Hitschfeld et al.}  \titlerunning {Gravitational
       stability of M51}
   \maketitle

\begin{figure*}[t]  
  \centering  
  \includegraphics[angle=-90,width=18cm]{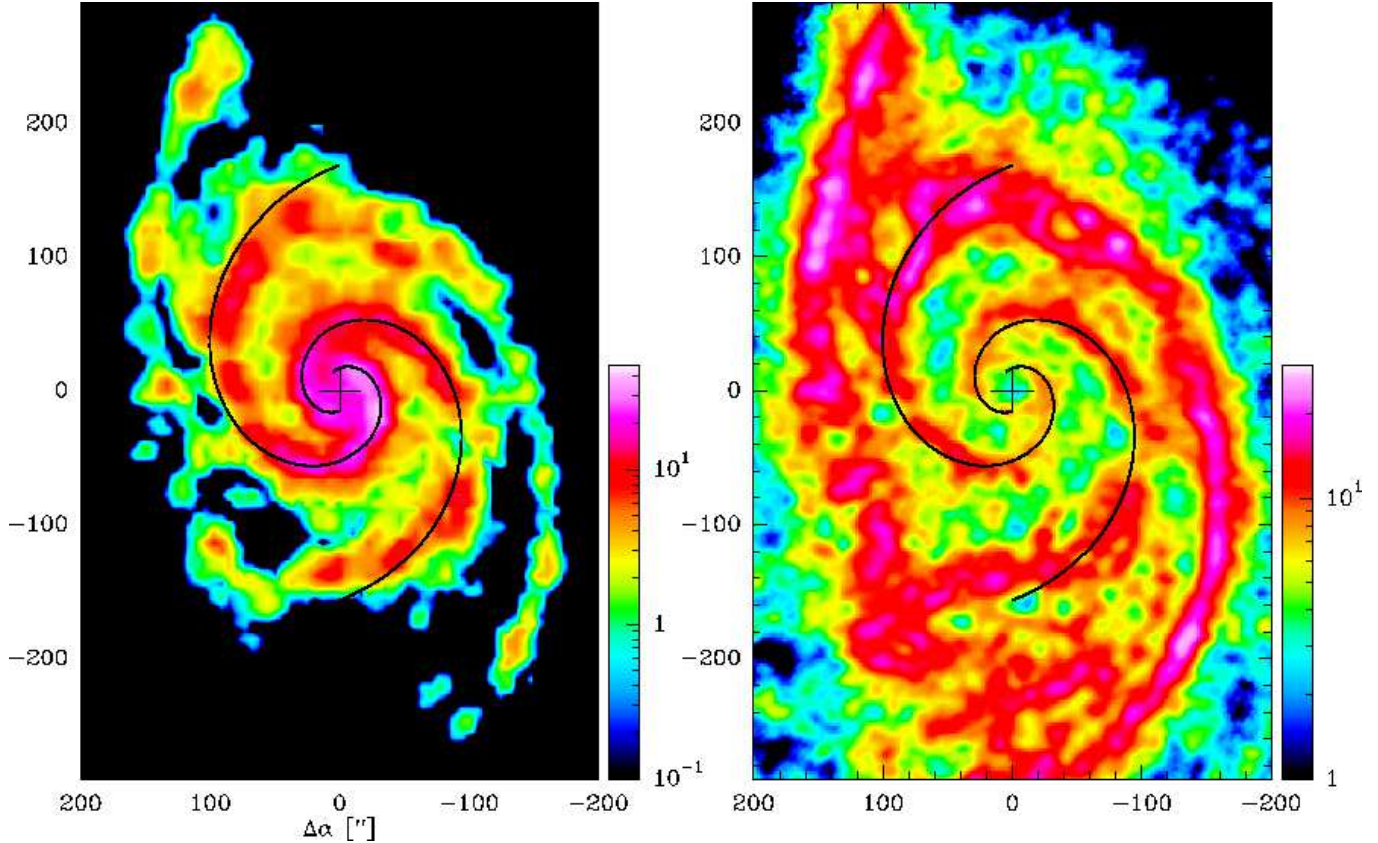} 
  \caption{{\bf Left: }Map of integrated CO 2--1 intensities on the antenna
    temperature scale in K\,km\,s$^{-1}$ at $11''$ resolution
    (cf.\,Paper\,I).  A logarithmic spiral is shown in all Figures of
    M51 to guide the eye. {\bf Right:} Map of \HI\ column densities \citep{Walter2008} in units of
    $10^{20}$\,cm$^{-2}$ at a resolution of $\sim6''$.}
 \label{fig-co}  
\end{figure*}  

% \begin{figure}[t]  
%   \centering  
%   \includegraphics[angle=-90,width=9cm]{../pics_tex/m51_co21.eps} 
%   \caption{Map of integrated CO 2--1 intensities on the antenna
%     temperature scale in Kkms$^{-1}$ at $11''$ resolution
%     (cf.\,Paper\,I).  A logarithmic spiral is shown in all Figures of
%     M51 to guide the eye.  }
%  \label{fig-co}  
% \end{figure}  
% 
% \begin{figure}[t]  
%   \centering  
%   \includegraphics[angle=-90,width=9cm]{../pics_tex/hiwalter_1E20.eps}
%   \caption{Map of \HI\ column densities \citep{Walter2008} in units of
%     $10^{20}$\,cm$^{-2}$ at a resolution of $\sim6''$.  }
%  \label{fig-hi}  
% \end{figure}  

\section{Introduction} %%%%%%%%%%%%%%%%%%%%%%%%%%%%%%%%%%%%%%%%%%%%%%%%%%%%%%%
 
%%CKr 17-Aug-2008: introduction completely revised:
 
To understand the evolution and appearance of galaxies it is crucial
to study the interplay between stars and the interstellar gas and
dust. Stars are made within the dense and cold interiors of molecular
clouds.  It is generally believed that molecular clouds form from the
densest regions of more widely distributed, diffuse atomic clouds.
During their lifetimes, stars seed the gas with heavy elements,
energy, and momentum, all of which strongly affects their parental
environment, the subsequent formation of new stars, and their host
galaxy. The photo-destruction of molecular clouds leads to their
dispersal and the formation of diffuse atomic clouds, thereby
recycling the material.  The processes governing the phase transition
from the cold neutral medium to dense, molecular gas, and back again,
are however still unclear
\citep{rosolowsky2007,knapen2006,Heitsch2008}. Studies combining maps
of CO, the traditional tracer of molecular gas, \HI, and the cold
dust will form the basis for any understanding of their interrelation.
The spatial distribution of the various galactic components is best
studied in nearby face-on galaxies, like M51, which lies at a distance
of 8.4\,Mpc \citep{feldmeier1997}.
 
Two empirical laws describe star formation in galaxies. First, stars
are observed to only form efficiently above a critical gas surface
density that appears to be determined by the Toomre
\citep{toomre1964} criterion for gravitational stability. This star
formation threshold has been observed in many galaxies
\citep[e.g.][]{martin_kennicutt2001}. Note, however the
exceptions discussed e.g. by \citet{wong_blitz2002}. Second, a tight relation has
been found between the star formation rate and the surface density of
the total gas, i.e. the molecular and atomic gas, the Schmidt law.  In
this paper, we will focus on the Toomre criterion.
 
In its original version it considers a dynamical stability analysis of
a thin, gaseous, differentially rotating, single-component disk
against axisymmetric gravitational perturbations. Observational
studies of the Toomre threshold have usually assumed a constant
velocity dispersion of the gas, while \citet{schaye2004} questions
this assumption, arguing that the drop of velocity dispersion
associated with the transition from the warm to a cold phase of the
interstellar medium causes the disk to become gravitationally
unstable. In addition, recent studies have shown that the stellar
component contributes significantly to the gravitational stability and
cannot be neglected \citep{Boissier2003,MacLow2007}.

In \citet[][hereafter Paper\,I]{Schuster2007}, we combined new CO 2--1
data of M51 obtained at the IRAM-30m telescope using HERA, with \HI,
and radio continuum data from the literature to discuss
radial profiles of the molecular and atomic gas mass surface densities
and star formation rates, averaged over azimuth as a function of
radius in the disk of M51. In the following, we will simply speak of
radial averages.
 
% A complete $^{12}$CO 2--1 map of M51 has been presented in Schuster et
% al. (2007, hereafter Paper\,I).  We found that the radial average of
% atomic to molecular gas rises by two orders of magnitude from the
% center to the outskirts at 11\,kpc radial distance and follows a power
% law with radius. This is similar to the results in other galaxies
% \citep[e.g.][]{wong_blitz2002}.  In Paper\,I, we also studied the
% gravitational stability of the gaseous disc of M51. The radial average
% of the Toomre parameter for the gaseous disc Q$_{\rm gas}$ is up to a
% factor of 5 above the critical threshold in the center while the outer
% parts are near gravitational instability, i.e.  near 1.
 
In the present paper, we combine the CO 2--1 data with \HI\ VLA data
from the THINGS survey \citep{Walter2005,Walter2008}, 850\,$\mu$m
SCUBA dust continuum data \citep{Meijerink2005}, and a $K$-Band 2MASS
image.  We use the CO and \HI\ data to derive maps of the mass surface
densities of the molecular and atomic gas, and maps of their velocity
dispersions. Maps of the total gas and of the ratio of molecular to
atomic gas are then discussed.  We separate the total gas mass surface
density map into an underlying exponential disk and the spiral arm
structure. The results are then compared with the dusty exponential
disk found by \citet{Meijerink2005} in their SCUBA map of M51. This
allows us to study the variation of the gas-to-dust ratios in the disk
and in the arms.
 
In preparation for the Toomre analysis, we discuss the velocity
dispersions of the molecular and atomic gas. Combining the data of the
molecular and atomic gas with the map of stellar mass surface density
derived from the 2MASS data, we then present the radial average and
a map of the total Toomre Q parameter describing the
gravitational stability of the disk.
 
% In the present study, we present the total gas density map and a
% molecular to atomic gas ratio map of M51 combining the CO map
% presented in Paper\,I with \HI\ VLA data from the THINGS survey
% \citep{Walter2005,Walter2008}. We study an underlying exponential
% distribution of the total gas density in the disc of M51 and compare
% our results with previous studies of an exponential dust disk found by
% \citet{Meijerink2005}. The $^{12}$CO 2--1 velocity field of M51, its
% deviations from purely rotational motions and the equivalent widths of
% the $^{12}$CO 2--1 emission are also presented. Taking advantage of
% this information, we compute the radial averages of the Toomre
% Q-parameters combining the contributions of the molecular-, atomic-,
% and stellar material.  We also present a map of the total Q-parameter.
 
%%CKr 17-Aug-2008. ----------------------------------------------

\section{Observations and Data}

Observations of the $^{12}$CO 2--1 emission from M51 were conducted
with the IRAM 30m telescope in February 2005. The half power beam
width (HPBW) is $11''$.  For detailed information of the observations
we refer to Paper\,I.

Additionally, we use data of the \HI\, Nearby Galaxy Survey (THINGS)
\citep{Walter2005,Walter2008}. M51 was observed in the VLA D, C, and B
array configurations and presented in \citet{Kennicutt2007}. In
D-configuration the interferometer is sensitive to scales of 15$'$
allowing to detect almost the total flux for M51. The angular
resolution in the integrated intensity image is 5.86$''$$\times$
5.56$''$. The sensitivity of the THINGS data is 0.44 mJy beam$^{-1}$ for 5.2
km/s channel width \citep{Walter2008}, which is about a factor
4 deeper than the VLA data of \citet{rots1990} used in Paper\,I. 
%
% \citet{rots1990} obtained 1.4 mJy beam$^{-1}$ for a similar beam size but
% 10.3 km/s channel width. This a factor of 4 lower sensitivity compared
% to that of the THINGS map at the same velocity resolution.

To derive stellar densities, we used the $K$-Band images of the Two Micron All
Sky Survey (2MASS) Large Galaxy Atlas \citep{Jarrett2003}.% at 3" resolution 

\section{Molecular and atomic gas}

\begin{figure}[t]  
  \centering  
  \includegraphics[angle=-90,width=9cm]{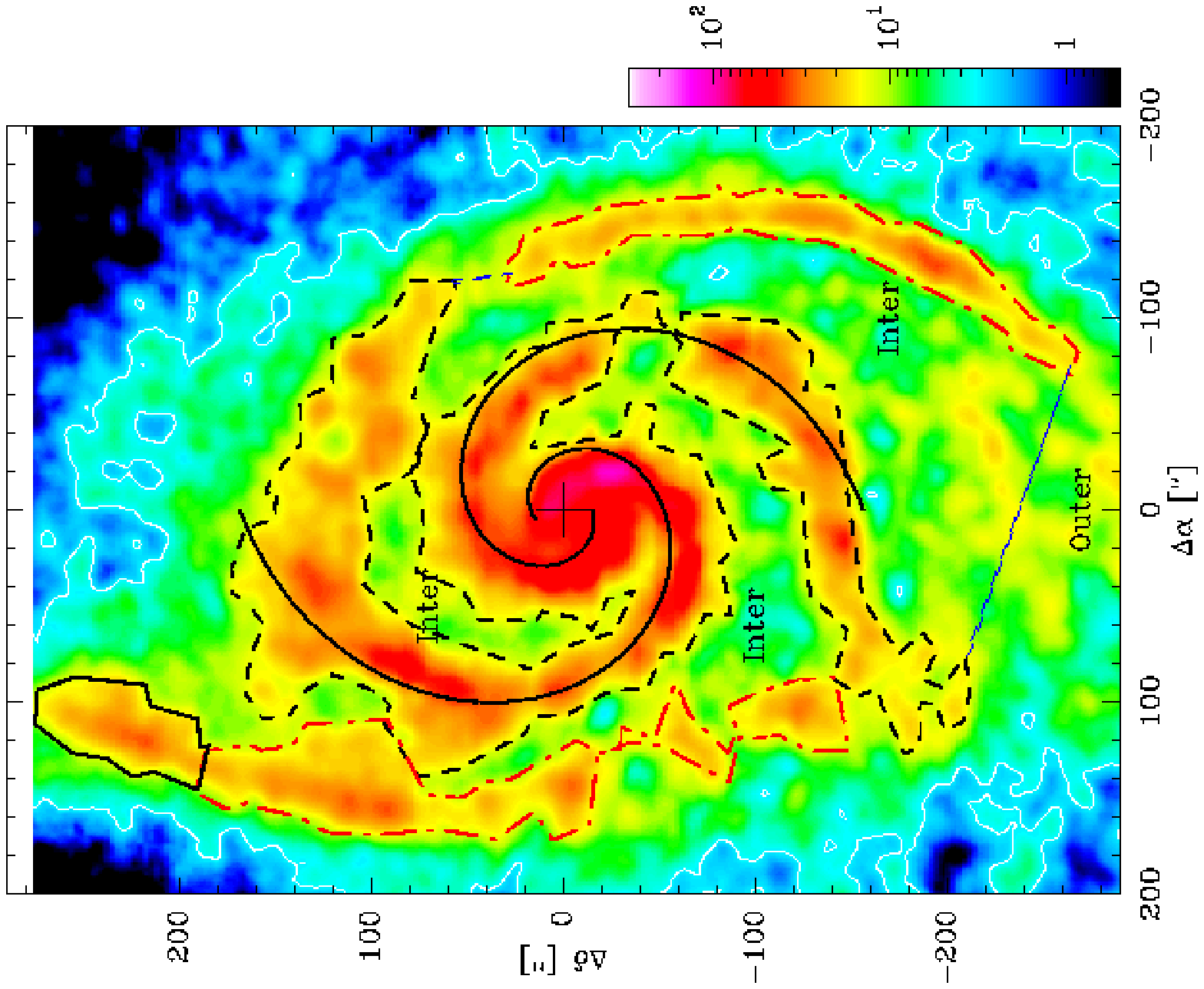}
  \caption{Map of the total gas mass surface density $\Sigma_{\rm
      gas}$ in M51 in \msun\,pc$^{-2}$ derived from the $^{12}$CO 2-1
    and \HI-emission. The cross marks the (0,0) center position at
    $\alpha=13\fh29\fm52\fs7$, $\delta=47\degr11\arcmin43\arcsec$
    (eq.2000). The 1-$\sigma$ value of the total gas density map is
    0.8 \msun\,pc$^{-2}$.  Contours mark regions discussed in \S
    \ref{totalgas_section} and Table \ref{totalgas_m51_properties}.
      }
 \label{total_gas}  
\end{figure}  

\subsection{The total gas surface density}
\label{totalgas_section}

In Fig.\ref{total_gas} we present a map of the total gas surface
density $\Sigma_{\rm gas}$. This map is a major ingredient of the
gravitational stability analysis we will discuss later in this paper.
% It combines the molecular and atomic local distribution in the disk of
% M51.

To convert from the observed integrated intensities to the total gas
surface densities, we use the relation $\Sigma_{\rm
  gas}=1.36\times(\Sigma_{\rm H2}+\Sigma_{\rm \HI}$) taking into
account the mass contribution from He (cf. Paper\,I).
% The molecular mass is estimated assuming a constant
% $^{12}$CO2-1/$^{12}$CO1-0 ratio of 0.8 and a quarter of the galactic
% X-factor $X_{\rm MW}=2.3\,10^{20}$cm$^{-2}$(Kkms$^{-1})^{-1}$
% \citep{Schuster2007,gb1993a,strong1988}.
%
The atomic surface density was calculated from the THINGS \HI\ column
density image of M51 (Fig.\ref{fig-co}), smoothed to 11$''$ resolution.  Two logarithmic
spirals (shown in black) serve as a guide to delineate the inner
spiral arms.  They are adopted from \citet{Shetty2007} who identified
them from a BIMA $^{12}$CO 1-0 map of M51.

The total gas surface density in Fig.\ref{total_gas} peaks at 127
\msun\,pc$^{-2}$. The inner spirals showing the densest peaks just
south-west of the center show total gas surface densities of 80-100
\msun\,pc$^{-2}$.

We subdivided the total gas map (Fig.\,\ref{total_gas}) into five
different regions: the inner and outer spiral arms, the interarm
regions, the outer disk, and the companion galaxy. All regions above a
threshold surface density of 20\,\msun\,pc$^{-2}$ were attributed to
the spiral arms. The outer border of the outer disk of M51 was defined
by the 3$\sigma$ level of $\sigma_{\rm gas}=2.4$\,\msun\,pc$^{-2}$
($\sigma(\Sigma_{\rm H2})$= 0.36 K\,km\,s$^{-1}$, $\sigma(\Sigma_{\rm \HI\
})$= 0.44 K\,km\,s$^{-1}$). The threshold for the arms leads to the
identification of spiral arms which correspond roughly with the arms
seen in interferometric CO maps \citet{Shetty2007}, and those seen in
H$\alpha$-, and in 20\,cm radio-continuum emission
\citep{Scoville2001,Patrikeev2006}. 

Table\,\ref{totalgas_m51_properties} lists the average total gas
density and the average H$_2$ over \HI\ density ratio for the five
subregions. The average of the inner spiral arm region,
is $\Sigma_{\rm gas}\sim$26.8 \msun\,pc$^{-2}$.
%
% The western inner arm shows a significant gap near (-80$''$,-80$''$)
% leading to the lower average density.

The outer spiral arms show less gas surface density, with an average
total gas density of 18.8\,\msun\,pc$^{-2}$, including the south
western spiral arm and the north eastern spiral arm. This is
approximately a factor of 1.5 weaker than the spiral arms in the
central region. Few high density regions on the outer spiral arms
reach $\Sigma_{\rm gas}$ of 40--50\,\msun\,pc$^{-2}$ e.g. to the
south-west at ($-120'',-200''$) and the north-east at ($140'',100''$).
This is signifcantly weaker compared to the peaks in the center.

\begin{center}   
\begin{table}[h*]
\caption[]{\label{totalgas_m51_properties}   
{\small Properties of $\Sigma_{\rm gas}$ and $\Sigma_{\rm
H2}$/$\Sigma_{\rm HI}$ in M51 averaged over the areas shown in
Fig.\ref{total_gas}.}}
\begin{tabular}{lrrrrr}   
\hline \hline   
                             & $\Sigma_{\rm gas}$  &
                             $\Sigma_{\rm H2}$/$\Sigma_{\rm HI}$ \\    
                             & [\msun\,pc$^{-2}$] &
                              \\    

\noalign{\smallskip} \hline \noalign{\smallskip}   
inner spiral arms & 26.8  & 2.3  \\               
outer spiral arms & 18.8  & 0.21  \\                   
interarm region & 9.4  & 0.71 \\                   
outer disk      & 6.4     &   -   \\
NGC\,5195 - companion & 20.2 & 0.4 \\
\hline               % OK    
ratio inner/outer spiral & 1.4 & 11.0 \\
inner arm/interarm       & 2.9 & 3.2 \\
outer arm/interarm       & 2.0 & 0.3  \\
%   
%%%%%%%%%%%%%%%%%%%%%%%%%%%%%%%%%%%%%%%%%%%%%%%%%%%%%%%%%%%%   
\noalign{\smallskip} \hline \noalign{\smallskip}   
\end{tabular}   
\end{table}   
\end{center} 
  
The interarm regions show surface densities of 5-15 \msun\,pc$^{-2}$.
The average value is 9.4 \msun\,pc$^{-2}$.  The averaged inner
arm/interarm contrast in the total gas surface density is 2.9
confirming the results of \citet{gb1993a} in the integrated CO 2--1
intensities. 

%% the errorbeam discussion is not very fruitful, so leave it out:
%%
% Note that the arm/interarm contrast is a lower limit as error beam
% contribution affects the integrated intensities. In interarm regions
% the pick-up of arm regions will tend to increase the intensity while
% in arm regions the situation is reversed. The IRAM 30m telescope has
% three error beam contributions with beam widths of about $125''$,
% $180''$ and $950''$ at 230\,GHz. \citep{greve1998} found relative
% power contributions of 10-20\%, 12\% and 26\%, and a main beam
% efficiency of 42\%.  At the time of the M51 observations in 2005, the
% beam efficiency had been improved to 52\%, which must have led to a
% reduction of the power in the error beams. We did not try to correct
% for these error beams.

There are several spur-like structures bridging the gap between the
inner spiral structure and the outer spiral arms e.g. in the
north-western area at ($-40'',80''$) at a surface density of 20-30
\msun\,pc$^{-2}$. Also, in the north-east at ($100'',100''$) there is
a strong bridge between inner and outer spiral arms with a total gas
surface density of 20--40 \msun\,pc$^{-2}$. The gaps or minima in
total gas density seen in the spiral arms e.g. at ($-80'',-40''$) and
at ($-120'',60''$) are interpreted as the signature of 4:1
ultra-harmonic resonances by \citet{Elmegreen1989} applying density
wave theory.

\begin{figure}[t]  
  \centering  
  \includegraphics[angle=-90,width=9cm]{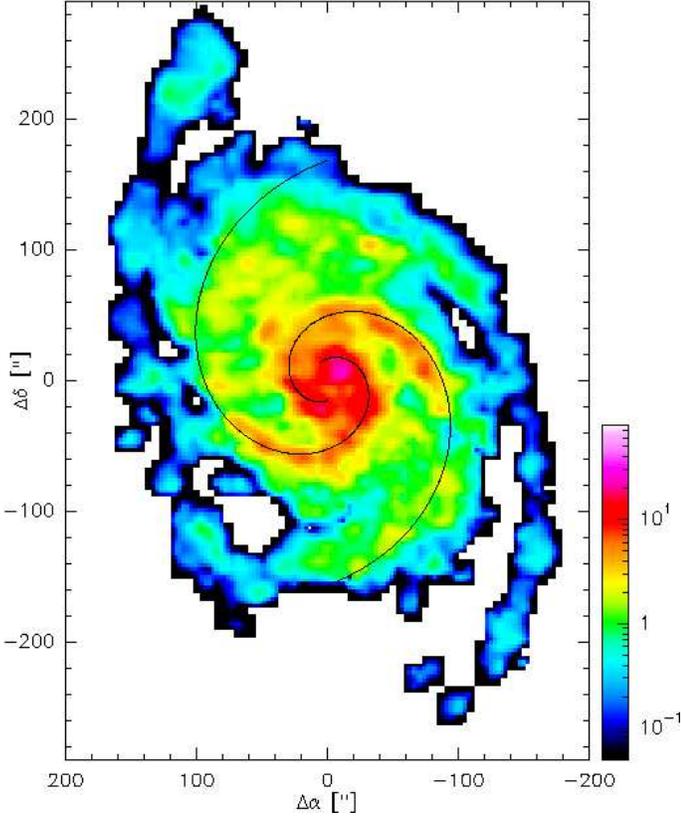}  
\caption{Map of the ratio of molecular- ($ \Sigma_{{\rm H}_2}$) and
atomic gas surface density ($\Sigma_{{\rm \HI}}$) in M51.
%
%derived from the $^{12}$CO 2-1 and \HI-emission. Regions without
%detection of $^{12}$CO 2-1 are blanked. The 1-$\sigma$ limit in the
%$^{12}$CO 2-1 data is 0.65Kkms$^{-1}$ corresponding to $\Sigma_{{\rm
%    H}_2}$= 0.36 \msun\,pc$^{-2}$.
%
Pixels below 2$\sigma$ of the masked-moment integrated intensity image
of $^{12}$CO 2--1 are blanked.  The 1-$\sigma$ limit of the $^{12}$CO
2-1 data is 0.65K\,km\,s$^{-1}$ corresponding to $\Sigma_{{\rm H}_2}$=
0.36 \msun\,pc$^{-2}$. As the \HI\ is more extended, the threshold of
the molecular gas limits the extend of the map. }
\label{ratio_hi_h2}  
\end{figure}  

\subsection{The  molecular gas fraction} 

Figure\,\ref{ratio_hi_h2} shows the ratio map of molecular versus
atomic gas surface density $\Sigma_{\rm H2}$/$\Sigma_{\rm \HI}$.
Overall, the map shows a steep decrease of the ratio from peak values
of around 50 in the center to 0.1 in the outer regions as already seen
in the radial average (cf. Paper\,I).  The spiral arms in the center
are the regions with the highest fraction of molecular gas. The
average ratio in the two inner spiral arms delineated in
Fig.\ref{total_gas} is 2.1 (Table\,\ref{totalgas_m51_properties}).
Outwards, at radii of more than about 100$''$, the ratio drops also in
the spiral arms. The north-eastern and south-western continuation of
the inner spiral arms show few regions with a ratio of 1 or slightly
above which are correlated with the regions of increased total gas
density (Fig.\ref{total_gas}).  The average ratio on the outer spiral
arms in the marked polygons is 0.21. This is a factor of 10 lower
compared to the inner spirals. The interarm region within the inner
$100''$ radius shows an average ratio of 0.71.  The molecular gas
surface density drops below the 2$\sigma$ threshold at larger radii
for most positions.

%  The dominant trend in the $\Sigma_{\rm H2}$/$\Sigma_{\rm \HI}$-ratio
%  is the decrease with galactocentric radius by at least one order of
%  magnitude. Additionally the $\Sigma_{\rm H2}$/$\Sigma_{\rm
%    \HI}$-ratio varies between spiral arm and interarm regions. {\bf
%    CHECK: these are repetitions of the above and not needed.} This
%  weaker effect, possibly induced by the spiral density wave, locally
%  influences the ratio. {\bf CHECK: why should that be ? This needs to
%    be substantiated or left out.} This enhances the molecular
%  fraction in the inner parts further and weakens the decrease {\bf
%    CHECK: that would mean that the molecular fraction in the outer
%    arms is larger than expected!?!?} of the molecular fraction in the
%  outer spiral arms.
%
%  {\bf CHECK: I suggest to rephrase as follows:} 
  The increase of the molecular-to-atomic gas fraction in the spiral
  arms may be triggered by the spiral wave increasing the hydrostatic
  pressure, and thereby the formation of molecular clouds from the
  atomic phases of the ISM.

Next, we will discuss the hydrostatic pressure in the disk of M51 as a
possible parameter governing the ratio of molecular and atomic gas.

\subsection{The hydrostatic pressure}

Recent investigations, e.g. by
\citet{Blitz2004ASP,Blitz2004,Blitz2006}, have shown a tight
correlation of the $\Sigma_{\rm H2}$/$\Sigma_{\rm \HI}$ ratio in
external galaxies and the hydrostatic pressure P$_{\rm hydro}$.
\citet{Elmegreen1993} theoretically studied the impact of hydrostatic
pressure and radiation field on the ratio of atomic and molecular gas.
He concludes that the pressure should be the dominant factor
determining this ratio. He predicts a power law $\frac{\Sigma_{\rm
    H_{2}}}{\Sigma_{\rm \HI}} \sim P_{hydro}^{2.2}$ of their radial
averages.  Following \citet{Blitz2004}, the hydrostatic pressure can
be estimated for an infinite disk with isothermal gas and stellar
layers from the midplane pressure in equilibrium:
\begin{equation}  
  P_{\rm hydro}= (2G)^{0.5}\Sigma_{\rm gas}\sigma_{\rm
    gas}(\rho_{*}^{0.5}+(\frac{\pi}{4}\rho_{\rm gas})^{0.5}).
\label{phydro}
\end{equation}  
Assuming the volume density of the gas $\rho_{\rm gas}$ is small
compared to the stellar density $\rho_{*}$ and a self-gravitating
stellar disk ($\Sigma_{*}=2\rho_{*}h_{*}$):
\begin{equation}  
  P_{\rm hydro}=0.84(G\Sigma_{*})^{0.5}\Sigma_{\rm gas}\frac{\sigma_{\rm gas}}{h_{*}^{0.5}}.
\label{phydro}
\end{equation}  

$\Sigma_{*}$ is the stellar surface density, which we will address in a later
section, $\Sigma_{\rm gas}$ the total gas density and $\sigma_{\rm gas}$ the velocity
dispersion of the gas, which we estimated from the $^{12}$CO 2--1 velocity
dispersion (cf. Paper\,I). For the stellar scale height, we adopt
h$_{*}=1$\,kpc, consistent with studies by \citet{Kruit1981,Kregel2002}.
$P_{\rm
 hydro}$ in equation \ref{phydro} is given in Nm$^{-2}$, assuming all
input parameters are in SI-units. 
% They find in their surveys of edge-on spiral galaxies a variation of h$_{*}$
% from 0.4-1.7\,kpc, with no dependence on the galacto centric radius.

\begin{figure}[h]  
  \centering  
  \includegraphics[angle=-90,width=8cm]{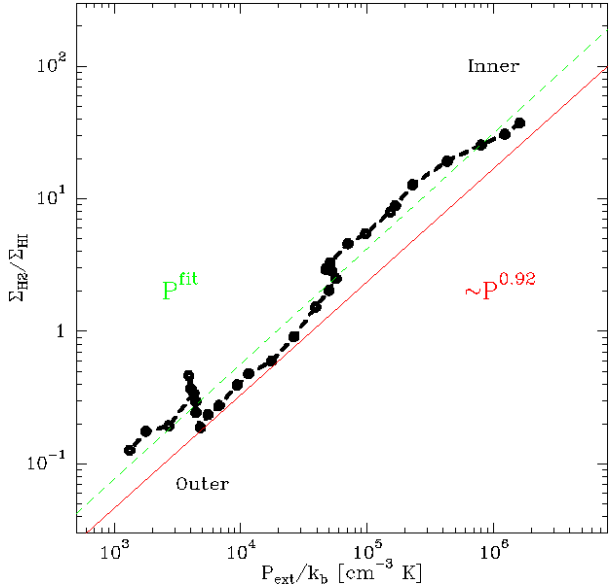}  
  \caption{The radial averaged fraction of molecular gas density $\Sigma_{\rm H_{2}}$
    and atomic gas density $\Sigma_{\rm \HI}$ is plotted against the
    hydrostatic pressure P$_{\rm hydro}$. The slope found in
    \citet{Blitz2006} is indicated in red. In green the linear
    regression fit to our data P$^{\rm fit}$ is shown. Going from left to
    right galactocentric radius decreases.}
 \label{ratio_gas}  
\end{figure}  

  As in Paper\,I, radial averages were created by averaging in
  elliptical annuli spaced by $10''$, including and taking into
  account points of no detection, and centered at the center position.
  The annuli are circular rings viewed at an inclination angle $i=
  20\degree$\, and with a line of nodes rotated from north to east by
  the position angle PA=170\degree.

  Figure\,\ref{ratio_gas} shows the $\Sigma_{\rm H_{2}}$/$\Sigma_{\rm
    \HI}$ ratio plotted versus the hydrostatic pressure P$_{\rm
    hydro}$, covering three orders of magnitude in pressure, the
  largest range measured so far in a galaxy disk. A linear regression
  fit yields a powerlaw:
\begin{equation}  
\frac{\Sigma_{\rm H_{2}}}{\Sigma_{\rm \HI}}=
(P_{\rm hydro}/P_{\rm 0})^{\alpha},
\end{equation}  
with the powerlaw coefficient $\alpha=0.87 \pm 0.03$ and $P_{\rm 0}$=
$(1.92\pm 0.07)\, 10^{4})$ cm$^{-3}$K. The slope of 0.87 is very
similar to the slope of 0.92 determined by \citet{Blitz2006} for a
sample of 14 spiral and dwarf galaxies, including M51. We find a
slightly lower $P_{\rm 0}$ compared to the mean result of
\citet{Blitz2006} in their sample with $(3.5\pm 0.6)\, 10^{4})$ cm$^{-3}$K.

This result underlines the importance of the hydrostatic pressure as
the physical parameter determining the fraction of molecular to atomic
material at a given radius on large scales. The nearly linear
dependence found over the whole pressure regime is expected by
theoretical predictions \citep{Blitz2006,Elmegreen1993}, if the
gravitational potential imposed by the gaseous component is small
compared to the stellar component. Note that the pressure dependence
of the molecular to atomic fraction is expected to break down when the
spatial resolution of the observational data is close to the
size of a single giant molecular cloud (GMC). A single GMC is expected
to have enhanced pressure due to self-gravity and the assumption of
pressure equilibrium is no longer valid \citep{Blitz2006}. 

\subsection{An exponential disk in M51}

\begin{figure}[t]  
  \centering  
  \includegraphics[angle=-90,width=9cm]{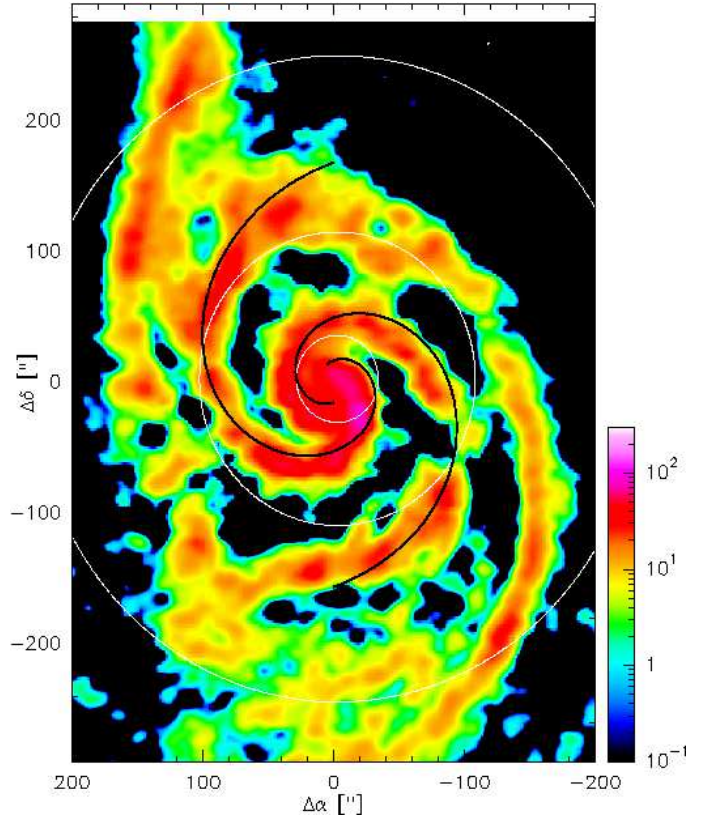}  
\caption{The residual map of the total gas surface density after removal of
  a fitted underlying exponential disk. The exponential disk is shown in
  contours with contour levels 1\msun\,pc$^{-2}$, 5\msun\,pc$^{-2}$ increasing
  to 15\msun\,pc$^{-2}$ in steps of 5\msun\,pc$^{-2}$.}  
 \label{res_exp}  
\end{figure}  

\begin{figure}[t]  
  \centering  
  \includegraphics[angle=-90,width=9cm]{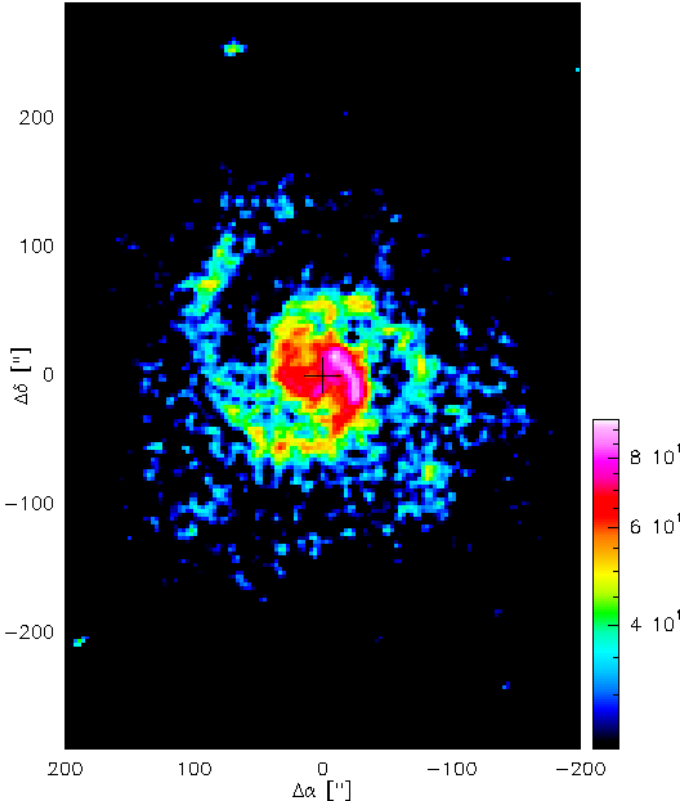}  
  \caption{The residual map of the 850$\mu$m dust emission after
    removal of a fitted underlying exponential disk
    \citep[Fig.2 in][]{Meijerink2005}. The 3$\sigma$-level is
    at 24\,Jy beam$^{-1}$. }
 \label{scuba_res_exp}  
\end{figure}  

\begin{figure}[h]  
  \centering  
  \includegraphics[angle=-90,width=9cm]{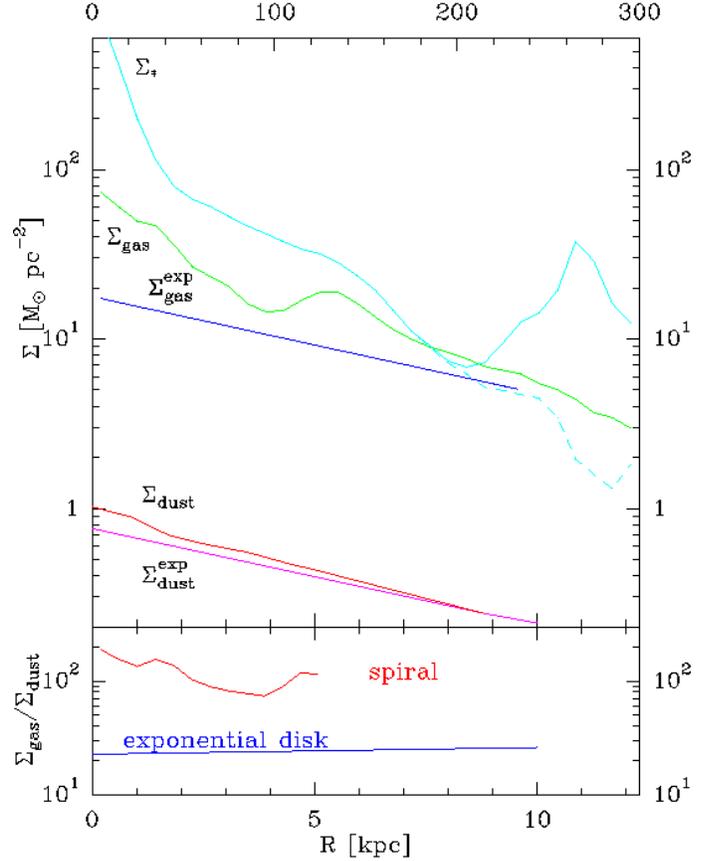}  
  \caption{{\bf (a)} Radial averages of surface densities. $\Sigma_*$
    is the stellar surface density. Drawn lines show averages
    including the companion galaxy. The dashed line shows stellar
    densities, ignoring the companion. $\Sigma_{\rm gas}$ is the
    surface density of the total gas. $\Sigma_{\rm gas}^{\rm exp}$ is
      the density of the fitted exponential gas disk. $\Sigma_{\rm
        dust}$ is the density of the dusty disk. $\Sigma_{\rm
        dust}^{\rm exp}$ is the density of the exponential dusty disk.
      Dust values below $3\sigma=24$\,Jy beam$^{-1}$ are ignored.
% dashed line: more precisely data above deltadelta 200'' are ignored.
%
   {\bf (b)} The gas-to-dust ratio for the exponential disk and for the residual data, 
   i.e. the spiral structure. }
 \label{radials}  
\end{figure}  

\begin{center}   
\begin{table}[h*]
  \caption[]{\label{expdisc_m51}   
    {\small Properties of the exponential disk of the total gas
derived in this paper compared to the properties of the exponential dust disk 
determined by \citet[][]{Meijerink2005}.}}
\begin{tabular}{lllrrr}   
\hline \hline   
\noalign{\smallskip} \noalign{\smallskip}   
% position as used in JCMT-CI and July-04-IRAM-30m observations:   
Total gas exponential disk: \\
$\Sigma_{\rm gas}$ scale length $h_{\rm gas}$ & 7.6 $\pm 0.1$ kpc \\
Amplitude: $\Sigma^{\rm peak}_{\rm gas}$  & 17.8 $\pm 0.2$ \msun\,pc$^{-2}$ \\
disk/total mass-fraction gas  & 61\% \\
% $\Sigma_{\rm HI}$ scale length h$_{\rm HI}$ & 8.7 $\pm 0.1$ kpc \\
\noalign{\smallskip} \hline \noalign{\smallskip}   
Dust exponential disk: \\
dust scale length $h_{\rm d}$ & 6.3 kpc \\
Amplitude: $\Sigma^{\rm peak}_{\rm dust}$ & 0.77 $\pm 0.2$ \msun\,pc$^{-2}$ \\
%Amplitude: $\Sigma^{\rm peak}_{\rm HI}$  & 12.4 $\pm 0.1$ \msun\,pc$^{-2}$ \\
disk/total mass-fraction dust & 55\% \\
%   
%%%%%%%%%%%%%%%%%%%%%%%%%%%%%%%%%%%%%%%%%%%%%%%%%%%%%%%%%%%%   
\noalign{\smallskip} \hline \noalign{\smallskip}   
\end{tabular}   
\end{table}   
\end{center}

Next, we investigate the model of an underlying exponential
distribution of the total gas density in the disk of M51 and 
compare it with the dust disk found by \citet{Meijerink2005}.
%\citet{Meijerink2005} observed 850$\mu$m dust emission in M51 at the
%JCMT using SCUBA and found an extended exponential cold dust disk,
%underlying the spiral arm structure. Here, we will derive the
%properties of the exponential disk using the two-dimensional total gas
%density distribution and compare to the properties of the dust disk.
We fitted an inclined elliptical exponential disk to the $\Sigma_{\rm
  gas}$ total gas surface density distribution map. We assumed a
centered disk at an inclination of 20\degree \, and a position angle
of 170\degree. The scale length $h_{\rm gas}$ and the amplitude
$\Sigma^{\rm peak}_{\rm gas}$ are determined from a 2-d fit of an
exponential distribution $\Sigma_{\rm gas}=\Sigma^{\rm peak}_{\rm gas}
\times$ exp($-R$/$h_{\rm gas}$). All pixels in the total gas density
map exceeding a threshold surface density of $\Sigma_{\rm
  gas}=20$\,\msun\,pc$^{-2}$ were blanked. This corresponds to the arm
regions we marked in Fig.\ref{total_gas} and discussed in \S
\ref{totalgas_section}. A significantly higher blanking threshold
leads to an overestimation of the exponential disk in the central
parts as spiral arm regions are included in the fitting.

\paragraph{Properties of the exponential disk}   

The scale length of the fitted exponential decay of the total gas disk
is $h_{\rm gas}= 7.6$ kpc (Table \ref{expdisc_m51}) with the peak at
17.8 \msun\,pc$^{-2}$.  The reduced $\chi ^{2}$ of the fit is 5.1. The
fraction of total gas mass contained in the exponential disk is 61\%.
Thus the spiral arms contain only $\sim40\%$ of the total gas mass.

\citet{Meijerink2005} fit a scale-length of the exponential dust disk
of $h_{\rm d}= 5.45$ kpc corresponding to $h_{\rm d}= 6.3$ kpc for a
distance of 8.4 Mpc, assumed in the present paper. Assuming an
isothermal disk, the fraction of dust contained in the exponential
disk is 55\%. 
Notably, the scale lengths of the total gas and of the dust exponential
disks are similar, as are the relative fractions of mass contained
in the disks.

Note that the shallow radial temperature profile
deduced by \citet{Meijerink2005} of 25\,K in the center to 17\,K in
the outskirts at 10\,kpc radial distance, only leads to a 35\% 
variation of corresponding dust surface densities if the temperature is
assumed to be constant. Thus, an isothermal dust
disk of 25\,K dust temperature and $\kappa_{850}= 1.2$cm$^{2}$\,g$^{-1}$ \citep{Meijerink2005} are
assumed in the calculation of the dust surface density. 
The dust surface density $\Sigma_{\rm dust}$ is 
\begin{equation} 
  \Sigma_{\rm dust} = M_{\rm dust}/A_{\rm beam} = 
  F_{850}\rm{d}^{2}/(\kappa_{850}B_{850}(T_{\rm dust})).
\end{equation}
$F_{850}= S_{850}/\Omega_{850} \times A_{\rm beam}$ is the integrated
flux. $S_{850}$ is the flux density, $\Omega_{850}$ the beam size of 15$''$ and
$A_{\rm beam}$ the beam area at the distance $d$=8.4 Mpc. $B_{850}$ labels the
Planck-Function at 850\,$\mu$m.

The radially averaged gas-to-dust mass ratio of the surface densities of
the exponential disk is shown in the lower box of
Figure\,\ref{radials}. The gas-to-dust mass ratio is nearly constant with
galacto-centric radius, reflecting the similar scale lengths. Values
vary only between 23 and 26, which is about a quarter of the canonical
Galactic gas-to-dust mass ratio of 100.  This constancy is in contrast for
example to the strong radial variation of the H$_2$/\HI\, surface
density ratio by more than a factor of 100.

The gas-to-dust mass ratio in the exponential disk is significantly lower compared to the
canonical Galactic value, whereas the metallicity of M51 is only slightly
supersolar with a shallow gradient with varying galactocentric radius \citep[e.g.][]{kramer2005}. 
However, several sources of uncertainty enter into this 
calculation which can not be resolved in the context of this study but might
drive the value towards the canonical gas-to-dust mass ratio of 100. In the inner
part H$_2$ dominates the total gas mass and is itself calculated via the
X-factor for M51 (see discussion in Paper\,I). The low gas-to-dust mass ratio in the
exponential disk, which constitutes the interarm regions, might
indicate an X-factor in the interarm regions closer to the Galactic X-factor (4 times the
value of M51). This will on the other hand not severely effect the gas-to-dust
mass ratio at larger radii where the atomic component is dominanting the molecular contribution.

Additionally, the assumption of a constant dust temperature $T_{\rm
  dust}=25$K and $\kappa_{850}= 1.2$cm$^{2}$\,g$^{-1}$ \citep{Meijerink2005}
might cause an overestimation of $\Sigma_{\rm dust}$ in the exponential disk.
A  4 times higher $\kappa_{850}$ yields a Galactic gas-to-dust mass ratio
value but seems unlikely as the values for
$\kappa_{850}$ determined in external galaxies are typically $\kappa_{850}=
0.7 - 2.1$cm$^{2}$\,g$^{-1}$ \citep{James2002,Alton2002}. A higher dust
temperature also drives the gas-to-dust mass ratio towards the Galactic value, with
a ratio of 100 for $T_{\rm dust}=121$K and $\kappa_{850}= 1.2$cm$^{2}$\,g$^{-1}$ as used above.
     
\subsection{Properties of the residual emission}   
Figures\,\ref{res_exp} and \ref{scuba_res_exp} show the residual total 
gas and dust maps, after subtracting the corresponding exponential disks.
The lower box in Figure\,\ref{radials} shows the corresponding radial
averages of the gas-to-dust mass ratio.

For the computation of dust surface density in the residual data, i.e.
in the spiral arms, we assume a uniform dust temperature $T_{\rm
  dust}=25$K and $\kappa_{850}= 1.2$cm$^{2}$\,g$^{-1}$
\citep{Meijerink2005} as above.

The resulting fraction of $\Sigma_{\rm gas}$ and $\Sigma_{\rm
  dust}^{\rm can}$ in the spiral arms, shown in Figure\,\ref{radials},
is not constant with radius, but rather varies by almost a factor 3
between about 70 and 190 within the first 5kpc where the emission is above the
3$\sigma$-threshold.

In contrast to the gas-to-dust fraction in
the exponential disk, we find a minimum of 73 at 3.8\,kpc radial
distand and a value of 186 in the center. Thus the gas-to-dust mass surface
density ratio is a factor of 3-8 higher in the spiral arms compared to
the underlying exponential disk.

As the exponential disk constitutes a minor
fraction of the gas mass in the spiral arms, it hardly mixes with the spiral
arm component.\\

The global gas-to-dust mass ratio, summing the mass components in the exponential
disk and the spiral arms in M51 upto the 3$\sigma$-limit of the dust
observations, is at 112 close the Galactic value.

\section{Stellar surface density}

\begin{figure}[h]  
  \centering  
  \includegraphics[angle=-90,width=9cm]{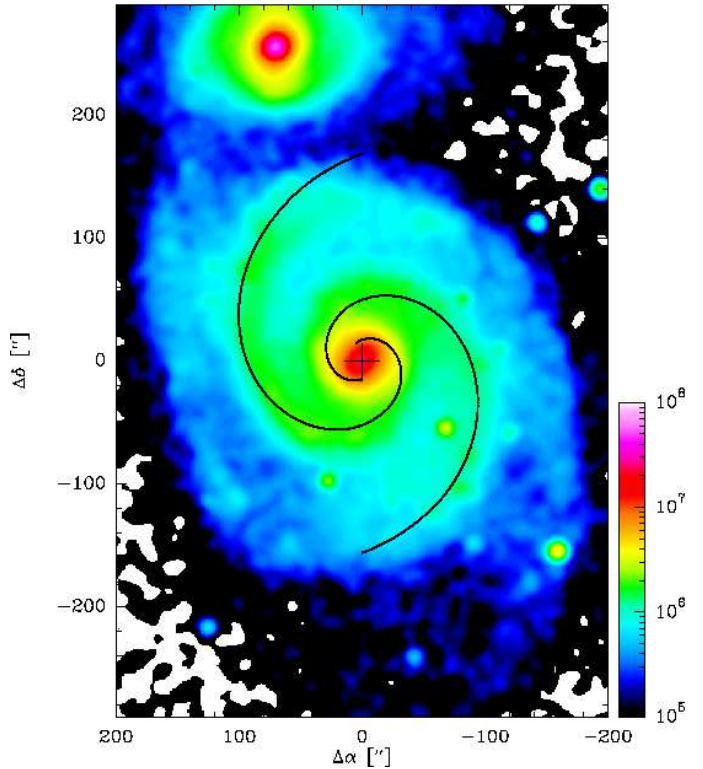}  
  \caption{Map of the $K$-band luminosity L$_{K}$ from the Two Micron
    All Sky Survey (2MASS) Large Galaxy Atlas \citep{Jarrett2003}
    smoothed to 11$''$ resolution. The units are \lsun.}
 \label{K_Band}  
\end{figure}  

The stellar surface density is an important parameter in determining
the stability of the disk. We use the $K$-Band images of the Two
Micron All Sky Survey (2MASS) Large Galaxy Atlas \citep{Jarrett2003}
to determine the stellar surface mass density $\Sigma_{*}$. The
$K-$Band is a reasonable tracer of the stellar mass density as it is
much less affected by extinction within M51 than e.g. the $B-$Band in
the optical.  The Galactic foreground extinction in the $K-$Band is
0.013 mag compared to 0.152 mag in the $B-$Band \citep{Schlegel1998}.

\citet{Bell2001} discuss mass-to-light ratios in the optical and
near-IR passbands.  Assuming an universal spiral galaxy initial mass
function (IMF), they find variations of the mass-to-light ratio of up
to 7 in the optical, 3 in the $B-$Band and 2 in the $K-$Band.  To
convert from luminosities to solar masses, we use a $K-$Band
mass-to-light ratio $M_{K}/L_{K} = 0.5$ \msun / \lsun
\citep{Bell2001}. The stellar surface density is then determined from
the $K$-Band luminosities via $\Sigma_{*}= 4.27\,10^{-5}\times L_{K}$
\msun\,pc$^{-2}$ for a distance of 8.4 Mpc to M51 and an angular
resolution of 3$''$.

The radial average of the stellar surface density $\Sigma_{*}$ is
shown in Fig.\,\ref{radials}. It decreases from a central value of
$\Sigma_{*}$=810\,\msun\,pc$^{-2}$ to 7\,\msun\,pc$^{-2}$\ before
increasing to 12 \msun\,pc$^{-2}$, due to the influence of the
neighboring galaxy NGC\,5195.

The fraction of stellar-to-gaseous mass density
$\Sigma_{*}/\Sigma_{\rm gas}$=10.5 is high in the center, decreasing
outwards to values below 1 at a radial distance of about 8\,kpc,
before increasing again due to NGC\,5195.

\section{Kinematics of the $^{12}$CO 2--1 data}   
\subsection{Velocity field and streaming motions}
 
\begin{figure*}[t]
  \centering
  \includegraphics[angle=-90,width=18cm]{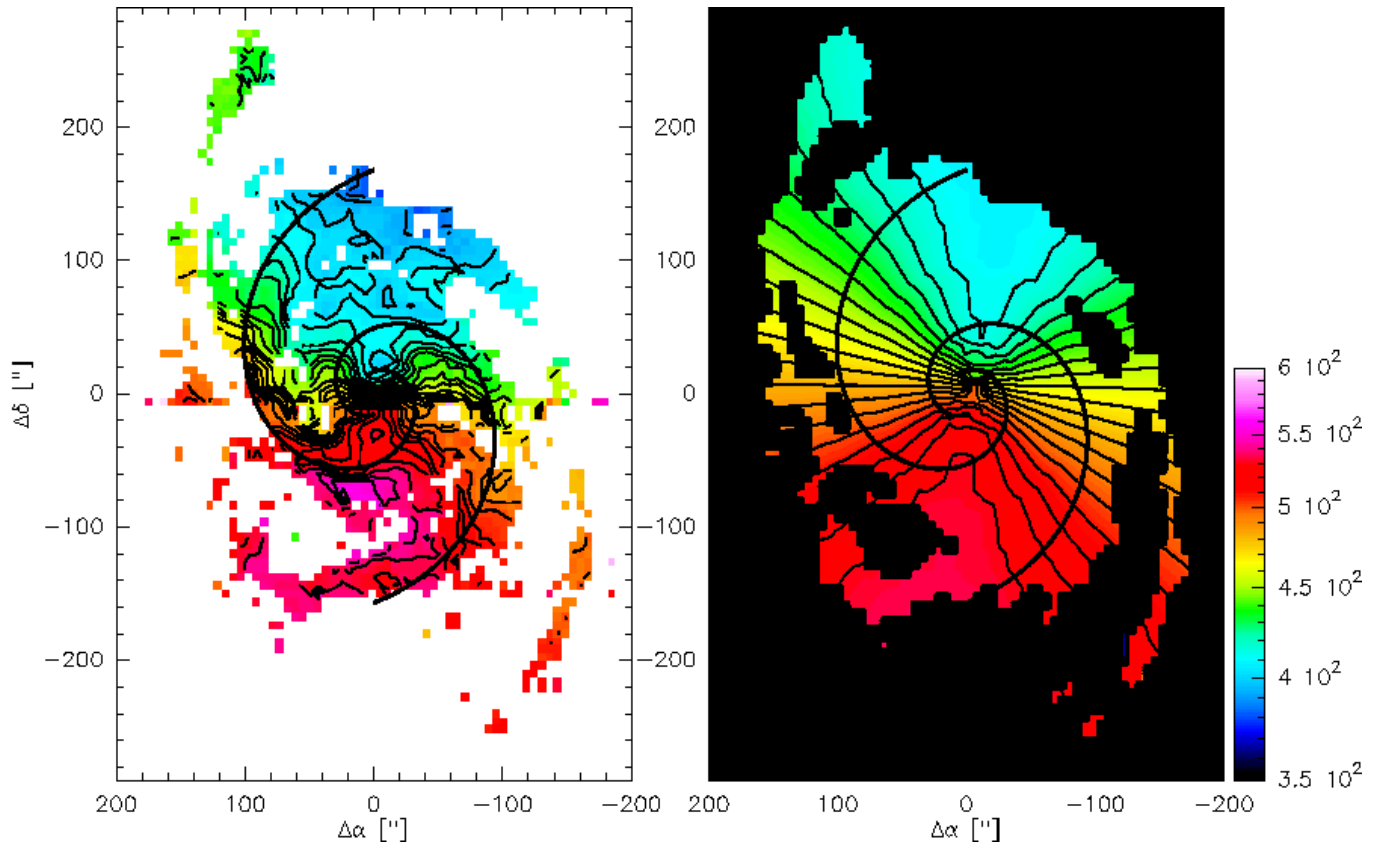}
  \caption{{\bf Left:} Map of the first moment $v_{\rm obs}=\int
    T\,v\,dv/\int T\,dv$ showing the intensity weighted mean velocity
    of the $^{12}$CO 2--1 data. Data below the 5-$\sigma$-level were
    blanked. {\bf Right:} Map of the calculated velocity field $v_{\rm
      mod}$ using the rotation curve of \citet{gb1993a,gb1993b}. The
    mean velocities range from 370 to 550\,km s$^{-1}$. Contour levels
    are in steps of 7km s$^{-1}$.}
 \label{velo_both}
\end{figure*}
 
Figure\,\ref{velo_both}a shows the observed velocity field of the
$^{12}$CO 2--1 data ranging from 370 to 550\,km s$^{-1}$ with the
northern part blue-shifted towards us and the southern part of the
galaxy red-shifted due to the inclination of the galaxy.
 
To calculate the velocity field for purely circular motions, we use
the rotation curve derived by \citet{gb1993a,gb1993b} and a 
position angle of $\Theta_{\rm MA}^{\rm mean}=170 \,\degree$.
In addition, we assume a constant
inclination of $i=20\degree$ and a central velocity of v$_{\rm
  sys}=$472 km s$^{-1}$ \citep{tully1974,Shetty2007}. The velocity
field at each point in the map is then given by
\begin{equation}
  v_{\rm mod}=v_{\rm sys}+v(R)\times\cos(\Theta-\Theta_{\rm MA}^{\rm mean})\times\sin(i).
\label{equ_velofield}
\end{equation}
with the azimuthal angle $\Theta$. The resulting map is presented in
Figure\,\ref{velo_both}b.
 
The two maps resemble well, i.e. the dominant component creating the
observed velocity field is the differential rotation curve $v(R)$
creating the striking X-shape of the iso-velocity contours.
 
\begin{figure}[h]
  \centering
%m51_residuals.eps aus die wurzel der quadrierten residuals -> nur positiv
  \includegraphics[angle=-90,width=9cm]{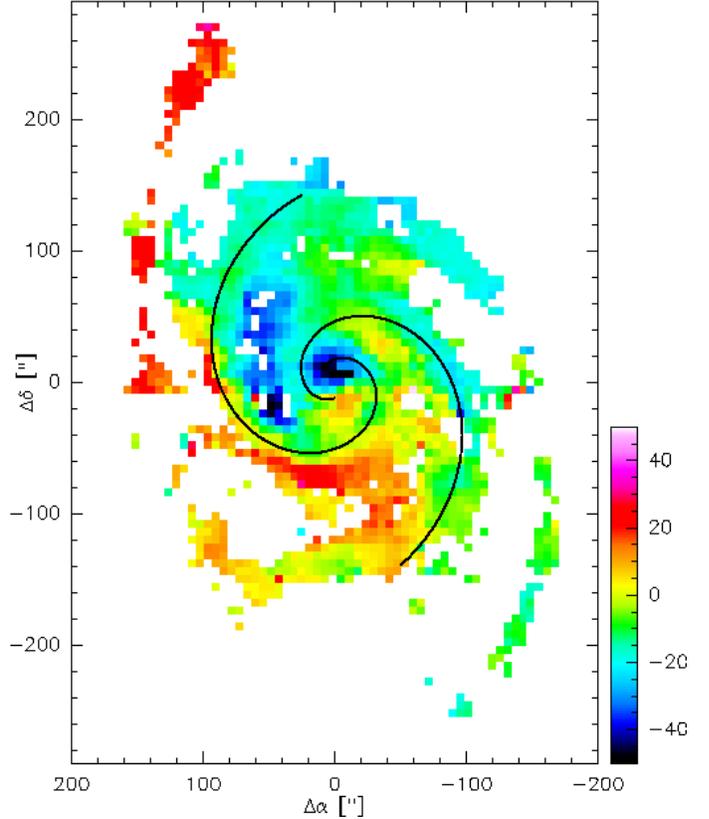}
  \caption{Map of non-circular streaming motions $\Delta v_{\rm res}$, i.e.
    the deviations of the observed velocity field from circular
    motions following the rotation curve. }
 \label{residual_sigma}
\end{figure}

However, the observed velocity field shows in addition many complex
structures and distortions not expected from purely rotational
motions.  We define the residual velocity as the difference of the
observed velocity field and purely rotational motions from the
rotation curve: $\Delta v_{\rm res}=v_{\rm obs}-v_{\rm mod}$.
Figure\,\ref{residual_sigma} shows a map of the residual velocities,
which we will call streaming motions henceforth.
 
Streaming motions vary between $\pm45$\,km s$^{-1}$. Strong residual
velocities are seen in the center and near the inner spiral arms, e.g.
near ($0''$,$-85"$) and ($60"$,$20''$). 
Both outer arms do not show significant streaming motions.  We
speculate that their locations at radii of $\sim 120-170 ''$ from the
galaxy center, lie close to the predicted location of the corotation
radius at $160 ''$ \citep{gb1993b,gb1993a}.  At this position no net
streaming is expected.
 
Here, we don't go into more detail, as streaming motions in M51 have
previously been discussed by \citet{gb1993b} using older 30m data and
by \citet{aalto1999} using OVRO data. In addition, \citet{kuno1997}
discussed the streaming motions in M51 based on $^{12}$CO 1--0 data observed
with the NRO 45m telescope. More recently, \citet
{Shetty2007} use BIMA $^{12}$CO 1--0 data of the entire galaxy at
$6''$ resolution to discuss radial and tangential streaming motions in
detail.  They determined the systemic velocity, position angle, and
inclination from a two-dimensional fit to the data.
 
\begin{figure*}[t]  
  \centering  
  \includegraphics[angle=-90,width=18cm]{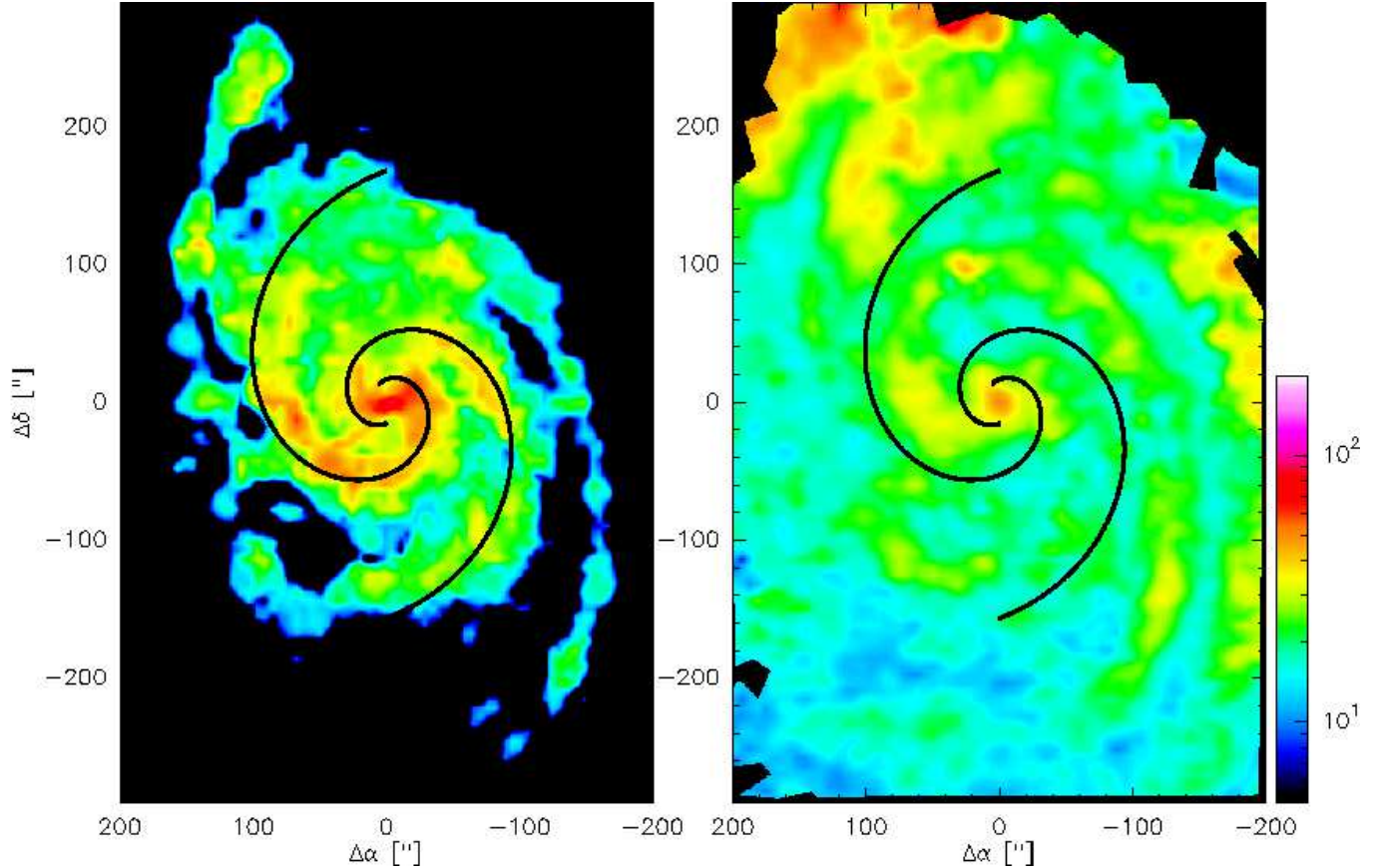}  
  \caption{Velocity dispersions of the molecular and atomic gas in
    M51. {(a)} map of observed widths of $^{12}$CO 2--1 in
    km s$^{-1}$.  
%
%    Contours show the velocity field from 350\,kms$^{-1}$ to
%    600\,kms$^{-1}$ in steps of 7\,kms$^{-1}$. 
    {\bf (b)} Velocity dispersion of \HI, i.e. map of the square root of
    the second moment in km s$^{-1}$. Positions where the \HI\ drops
    below the 3$\sigma$ level of 1.33\,\msun\,pc$^{-2}$ are blanked.
}
\label{equ_width}  
\end{figure*}  

\begin{figure}[h]  
  \centering  
  \includegraphics[angle=-90,width=7cm]{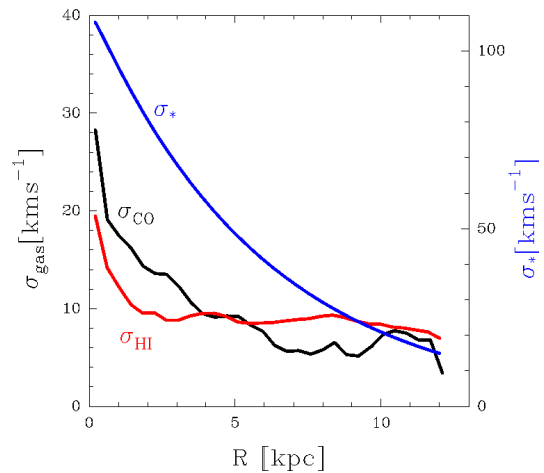}  
  \caption{Radial averages of the velocity dispersions $\sigma={\rm
      FWHM}/(2\sqrt{2\ln(2)})$ of the molecular gas traced by CO, \HI,
    and stars.  }
 \label{radial_velo}  
\end{figure}  

\begin{center}   
\begin{table}[h*]
\caption[]{\label{equivwidths_m51_properties}   
{\small Equivalent widths $\Delta$v$_{\rm eq}$ of CO and \HI\,  in M51 averaged over the
  areas shown in Fig.\ref{total_gas}.}}   
\begin{tabular}{lrrr}   
\hline \hline   
                             & $\Delta$v$_{\rm eq} (\rm CO)$ & $\Delta$v$_{\rm
                               eq}(\HI\ )$ \\    
                             
                             & [km s$^{-1}$] & [km s$^{-1}$] \\    

\noalign{\smallskip} \hline \noalign{\smallskip}   
inner spiral arms & 25.9 & 21.5 \\               
outer spiral arms & 11.1 & 18.3 \\                   
interarm region & 7.3 & 19.4 \\                   
\hline               
ratio inner/outer spiral & 2.3 & 1.2 \\
inner arm/interarm       & 3.6 & 1.1 \\
outer arm/interarm       & 1.5 & 0.9 \\
%   
%%%%%%%%%%%%%%%%%%%%%%%%%%%%%%%%%%%%%%%%%%%%%%%%%%%%%%%%%%%%   
\noalign{\smallskip} \hline \noalign{\smallskip}   
\end{tabular}   
\end{table}   
\end{center}

\subsection{Velocity dispersion of the molecular and the atomic gas}

The velocity dispersion of the gas is a signature of the turbulent
interstellar medium and systematic motions within the beam.  The
turbulent medium and the differential rotation curve hinders the
medium to collapse under its gravitational pull, as described by the
Toomre criterion.

% gas is locally stabilized against collapse For the analysis of the
% gravitational stability using the Toomre criterion, we need the
% velocity dispersion of the gas as an important parameter possibly
% stabilizing the gas locally.

\subsubsection{ $^{12}$CO 2--1}

Figure\,\ref{equ_width}a shows a map of equivalent $^{12}$CO 2--1 line
widths $\Delta v_{\rm eq}=\int T dv / T_{\rm pk}$ measuring the
velocity dispersion.\footnote{For a Gaussian line profile, the equivalent width
corresponds to the full width at half maximum (FWHM).}

We will refer to $\Delta v_{\rm eq}$ in the following as $\Delta v_{\rm obs}$ the
observed line width.

The broadest widths are observed in the center, reaching up to
100\,km s$^{-1}$. In the disk of M51 spiral arms show enhanced line widths
relative to the interarm regions. The inner spiral shows strongly
enhanced widths relative to the inter arm molecular gas.  Along the
inner spiral, the widths decrease from about 70 to 20\,km s$^{-1}$. At
the same time, the contrast between the arm and interarm dispersion
diminishes. Near (60",100"), arm and interarm molecular gas shows the
same dispersion of about 15-20\,km s$^{-1}$. Overall, regions with
distances greater than aboutg 100$''$ from the center show
significantly lower widths of 15 to 20\,km s$^{-1}$.  This drop of CO
line widths with radius is also seen in a plot of radial averages
(Fig.\,\ref{radial_velo}) and in the table of line widths averaged
over the inner, the outer, and the interarm regions
(Table\,\ref{equivwidths_m51_properties}).  The outer south-western
spiral arm shows uniform widths of around 15\,km s$^{-1}$.
\paragraph{Influence of systemic motions}
The $^{12}$CO 2--1 line widths map in Figure\,\ref{equ_width}a shows no strong
systematic broadening of the observed line widths $\Delta v_{\rm obs}$ with varying angle in the
disk which would indicate that a major fraction of the observed line widths
originates in the galactic rotation of M51 (modeled in
Figure\,\ref{velo_both}b) polluting the observed line widths $\Delta v_{\rm obs}$. 
The velocity gradient within the observing beam (FWHM=11$''$) due to the velocity field will
broaden the observed line widths and thus lead to an overestimation of the
intrinsic line width $\Delta v_{\rm obs}$. In the following, we will quantitatively 
estimate the magnitude of this effect on the observed line widths
at some selected positions in the galaxy disk where velocity gradients are
suspected to be the  highest due to projection: the inner region (0$''$/0$''$)
and along the minor axis. Elsewhere in the disk the gradient of the velocity field is
significantly lower (Figure\,\ref{velo_both}b) and additionally the observed line widths are similar to
the outer parts of the minor axis (Figure\,\ref{equ_width}a).   

\begin{figure}[h]  
  \centering  
  \includegraphics[angle=-90,width=8cm]{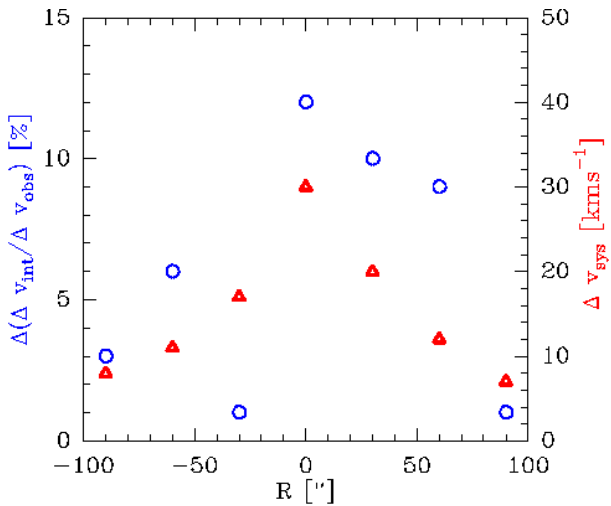}  
  \caption{The overestimation of the intrinsic line widths by the observed line widths near the minor
    axis $\Delta (\Delta v_{\rm int}/\Delta v_{\rm obs})$ in blue circles and the systemic
    velocity gradient in the beam $\Delta v_{\rm sys}$ in red triangles for the $^{12}$CO 2--1 data.  
}
\label{fig_vsys}  
\end{figure}

\citet{Sakamoto1996} corrected for the broadening due to systemic motions
$\Delta v_{\rm sys}$ using: $\Delta v_{\rm int}^{2}=\Delta v_{\rm obs}^{2}-\Delta v_{\rm sys}^{2}$.
$\Delta v_{\rm sys}$ is the velocity variance across the beam as caused by the modeled
large scale velocity field. We apply this method
to quantify the influence of the rotational motion at the central position
(0$''$,0$''$) and six other positions near the minor axis of the galaxy from the
center outwards (Figure\,\ref{fig_vsys}).       

The relative error of $\Delta v_{\rm obs}$ is $12\%$ at the center and
decreases outwards along the minor axis with values of less than
$3\%$ at distances $90''$ away from the center. This relative error is shown
together with $\Delta v_{\rm sys}$ for each of the six studied positions in
Figure\,\ref{fig_vsys}.
We estimate the effect of the polluted $\Delta v_{\rm obs}$ on  $Q_{\rm tot}$ (next
section) using Gaussian error propagation for a reasonable range of values.
% Q$_{\rm \HI}=1-5$, Q$_{\rm H_{2}}= 1-10$ and Q$_{\rm *}= 5-20$.
The absolute change of $Q_{\rm tot}$ for an overestimation of the line
widths of 10$\%$ for the gaseous components (the broadening of stellar dispersion due to systemic
motions is negligible) results in an maximum absolute uncertainty of 0.16.
As this effect is minor, we will in the following neglect the broadening due to
rotational motions.    
\subsubsection{\HI}
Although line center of CO and \HI\, emission usually coincide quite well,
dispersions can be quite different.
Figure\,\ref{equ_width}b shows the square root of the second moment
M$_{2}^{1/2}= \sqrt{T\,v^{2}\,dv/\int T\,dv}$ of the \HI\ emission as
a measure of its velocity dispersion. For a Gaussian line profile,
M$_{2}^{1/2}$ corresponds to the FWHM. 

The \HI\ line widths show values of around 30--50\,km s$^{-1}$ in the
central part. In general, the spiral arms do not show up prominently
and the interarm regions show a larger dispersion than the spiral
arms, in strong contrast to the corresponding CO map. For example, the
inner interarm region near (0$''$,100$''$) shows velocity widths
reaching values of up to 40\,km s$^{-1}$ which is significantly higher
compared to the arm regions. Towards the companion galaxy, we see
enhanced widths, comparable to those in the central region. \HI\ shows
a strong north-south asymmetry with significantly lower line widths in
the souther parts of the galaxy.  This is less pronounced in the
CO-data. Several of these observations are reflected in the plot of
radial averages (Fig.\,\ref{radial_velo}) and in the table of line
widths averaged over the different regions
(Table\,\ref{equivwidths_m51_properties}).

In the plot of radially averaged dispersions
(Fig.\,\ref{radial_velo}), the dispersion of CO gas exceeds the
dispersion of \HI\ for radii up to 4\,kpc. The decrease of the
dispersion of the atomic component in the inner part from around 20 to
10 km s$^{-1}$ within 2\,kpc is steeper compared to the decrease of the
CO dispersion. The radial averaged atomic component then stays
constant at around 10\,km s$^{-1}$ out to radii of 12\,kpc. The CO
dispersion drops to 6\,km s$^{-1}$ below the dispersion of \HI\ at
radii of around 6\,kpc and increases to 11\,km s$^{-1}$ outwards due to
the increased dispersion in the center of the companion.

Our findings are consistent with the following scenario
\citep[cf.][]{gb1993a,rand1992}.  The dispersion of the molecular gas
is enhanced in the spiral arms due to frequent collisions of molecular
clouds which may be enhanced by streaming motions driven by the
density wave. The atomic clouds in the arms appear to form a different
population as their local velocity dispersion is much less compared to
the molecular gas.  We speculate that thus \HI\ in the arms stems
mainly from photodissociated H$_{2}$ in GMCs, where the dispersion has
already decreased due to dissipation. The interpretation of VLA \HI\
and OVRO CO observations by \citet{rand1992} also favors this
dissociation scenario.

The plot of radially averaged dispersions (Fig.\,\ref{radial_velo})
also shows the dispersion of disk stars. This is an important
parameter for determining the gravitational stability of the disk and
will be used in the following section.  To estimate the stellar
velocity dispersion in M51, we follow \citet{Boissier2003} and
\citet{Bottema1993}. They showed that the velocity dispersion follows
an exponential fall-off $\sigma_{\rm stars}=\sigma_{0}\, \rm{exp}(-R/
2\rm{h}_{\rm B})$ depending on the scale length of the B-band
$\rm{h}_{B}$ of the disk. The B-band scale length is $\rm{h}_{B}=2.82
kpc$ \citep{Trewhella2000} and the central stellar velocity dispersion
in M51 is $\sigma_{0}=113$ km s$^{-1}$ \citep{McElroy1995}. At radii upto 10
kpc, the velocity dispersion of the stars is larger than that of the
molecular and atomic gas components.

% \begin{figure}[h]  
%   \centering  
%   \includegraphics[angle=-90,width=9cm]{../pics_tex/m51_corr_disp.eps}  
% \caption{Map of the velocity dispersion of the molecular gas corrected for by
%   the residuals of modeled and observed velocity field.}  
%  \label{cleaned_sigma}  
% \end{figure}  
% \begin{bf}{CHECK: Discuss maps in more detail.}\end{bf}

\section{Gravitational stability}

\label{section_toomre}

\subsection{The combined Toomre parameter $Q_{\rm total}$}
We investigate the gravitational stability of the disk of M51
using the Toomre Q-parameter, taking into account the stellar and
gaseous contribution. In general, the Toomre stability criterion is
depending on the epicyclic frequency $\kappa(R)$, the velocity
dispersion $\sigma(R)$ of the component considered, and the surface
density $\Sigma$ \citep{toomre1964}, neglecting e.g. the influence of
magnetic fields:

\begin{equation}
Q = \frac{\kappa(R) \sigma(R)}{\pi \rm{G} \Sigma}.
\label{equn_toomre}  
\end{equation}  

The gas should collaps if Q drops below 1, when gravitation dominates
over dispersion and the epiclyclic frequency.

We treat the molecular gas, the atomic gas, and the stars in the disk
of M51 as three independent isothermal fluids using the expansion in
wavenumber by \citet{Wang1994} and neglecting higher order terms in
the wavenumber i.e. in the velocity dispersions.  The model assumption
is that the stars move through the gaseous medium without interaction.
For this first order analysis, we assume in addition that the dense
molecular clouds move frictionless through the diffuse atomic gas.
The total Toomre parameter then is:

\begin{equation}  
Q_{\rm tot}^{-1} = Q_{\rm H_{2}}^{-1}+Q_{\HI\ }^{-1}+Q_{*}^{-1}.  
\label{eq_sigma_crit}
\end{equation}  

For $\kappa(R)$, we use the rotation curve derived from CO, assuming
that it holds to the \HI\ gas and for the stars in the disk. For the
velocity dispersion $\sigma(R)$, we use the dispersions derived above
from CO, \HI, and the stars, respectively.

%The Q-parameter for a pure stellar disk is calculated analog to
%Q$_{\rm gas}$. 
%\begin{equation}  
%Q_{\rm tot}^{-1} = Q_{\rm gas}^{-1}+Q_{*}^{-1}.  
%\end{equation}  

\begin{figure}[h]  
  \centering  
  \includegraphics[angle=-90,width=7cm]{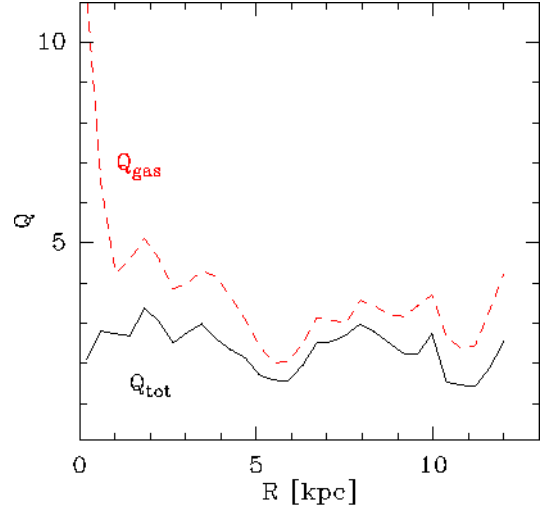}  
\caption{Radial average of the total Toomre parameter Q$_{\rm tot}$
including the stellar contribution. In red the radial average of the
gas-only Toomre parameter Q$_{\rm gas}$ is shown.
 % In blue the radial average of Q$^{\sigma}_{\rm total}$
 %   for a constant velocity dispersion of the gaseous component
 %   $\sigma=6$km s$^{-1}$ is shown and
}
 \label{radial_Q}  
\end{figure}

To calculate $Q_{*}$, we used the 2Mass $K$-band image smoothed to
11$''$, as described before. In Fig.\ref{radial_Q}, the radial
averages of Q$_{\rm tot}$ and Q$_{\rm gas}=(Q_{\rm H_{2}}^{-1}+Q_{\HI\
}^{-1})^{-1}$ are shown.

Taking into account only the \HI\ and CO gas, Q$_{\rm gas}$ drops from
values of more than 10 in the center to values of between 2 and 5 for
radii beyond about 1\,kpc, as has already been discussed in Paper\,I.
In contrast to Q$_{\rm gas}$, Q$_{\rm tot}$ is smaller and variies
only little between 1.5 and 3, at all radii, staying always close but
slighty above the critical value for gravitational instability of 1.
This finding indicates a self-regulating disk, keeping the total
Toomre Q parameter near 1.  We will enter into the discussion of our
results in the next section.

The importance of the stellar contribution has also been studied by
\citet{Boissier2003}, who presented radial averages of the Toomre
Q-parameter for their sample of galaxies, comparing $Q_{\rm gas}$ and
$Q_{\rm tot}$.  They assume a constant velocity dispersion and find
that the stellar contribution lowers the Q-parameter by up to 50\%
towards the critical threshold. \citet{Leroy2008} very recently also studied
the radial averages of the Q-parameter including the stellar contribution in a
sample of nearby galaxies from THINGS. They find a
marginally stable Q$_{\rm tot}$ in their sample of 23 nearby dwarf and spiral
galaxies in accordance with our M51 Q$_{\rm tot}$-results.

\subsection{The influence of streaming motions}

Below, we will show that the contribution of non-circular motions to
the measured velocity dispersions is small. The influence of streaming
motions on the local gravitational stability can therefore be ignored.

To test the influence of non-circular streaming motions on the
stability, we correct in the following the observed velocity
dispersion for these residual velocities $\Delta v_{\rm res}$, assuming
Gaussian line profiles:

\begin{equation}  
  \Delta v_{\rm corr}^{2} = 
  \Delta v_{\rm eq}^{2} -(v_{\rm obs}- v_{\rm mod})^{2}.
\end{equation}  

% $\Delta v_{\rm corr}$ is the intrinsic linewidth $\Delta v_{\rm eq}$,
% corrected for the broadening caused by streaming motions. 
%
This reduced dispersion leads to a reduction of Q$_{\rm H_{2}}$,
assuming that the non-circular streaming motions do not stabilize the
gas against collapse.

A local change of the observed line widths $\Delta v_{\rm obs}$ propagates
into Q$_{\rm gas}$ according to Gaussian error propagation. Typical
values of Q$_{\rm \HI}= 1-5$, Q$_{\rm H_{2}}= 1-10$ , $\sigma_{\rm
  CO}=10$km s$^{-1}$ and a variation of 50\% of the velocity
dispersion, yields absolute variations of Q$_{\rm gas}$ of upto 56\%
only.
% 0.04--0.56.

In the radial averages, the maximum relative change between the
corrected and uncorrected Q$_{\rm tot}$ is only 7\%. In the following,
we therefore neglect the streaming motions in the Toomre analysis, and
use directly the observed velocity dispersion for the calculation of
the Q-parameter.

\begin{figure}[h]  
  \centering  
  \includegraphics[angle=-90,width=9cm]{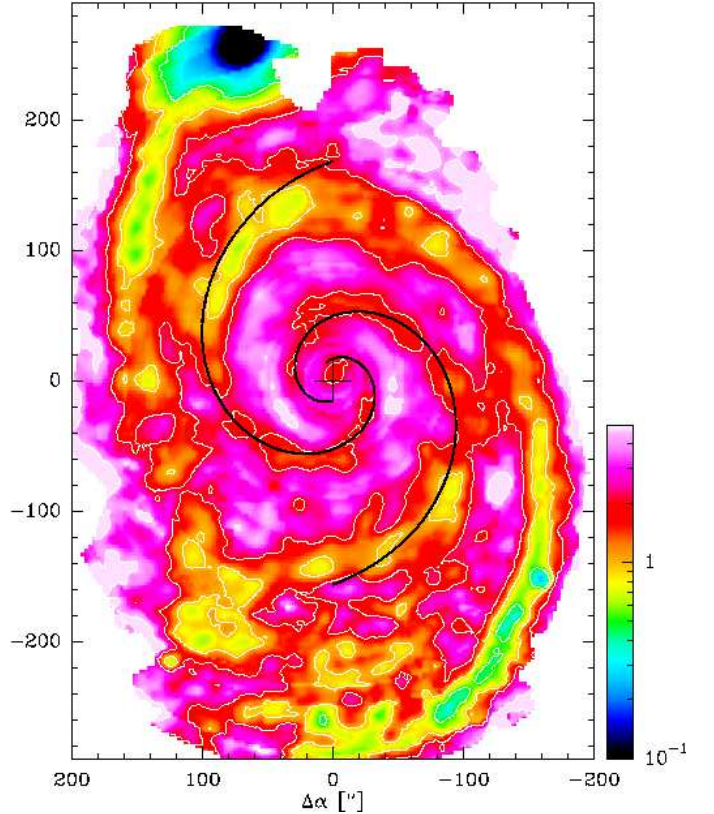}  
  \caption{Map of the total Toomre parameter Q$_{\rm tot}$ including
    the stellar contribution. We blanked pixels with $\Sigma_{\rm
      gas}$ below the 3$\sigma$-level. The contours show Q$_{\rm
      tot}$= 0.5, 1, 2, and 5.  }
 \label{toomre_Q_corr}  
\end{figure}  

\subsection{The local distribution of Q$_{\rm tot}$}
The local values of the Toomre Q-parameter Q$_{\rm tot}$ at each
position are studied in the following. Numerical simulations show that the local Q values
can deviate significantly from the radially averaged values. \citet{Wada1999} use two-
dimensional hydrodynamical simulations to study the gravitational stability of
the central region of a galactic disk. While global Q-parameter values of
their model disk are significantly above the threshold for gravitational
collapse, locally the disk shows unstable regions spatially correlated with cold and
dense clumps.   

Figure\,\ref{toomre_Q_corr} shows a map of the Toomre Q-parameter
Q$_{\rm tot}$ in M51. In general, spiral arms show smaller values of Q$_{\rm
  tot}$, i.e. they are more prone to collapse under the gravitational
pull, than the surrounding interarm gas. This is expected as star
formation occurs predominantly in the spiral arms. 

The inner spiral arms show $Q_{\rm tot}$ values between 1 and 2 near
to the threshold for gravitational stability.  The regions between the
inner spiral arms show values up to a factor of 2 to 3 higher as they
lack surface density.  The outer south-western spiral arm is almost
everywhere critical to gravitational collapse with $Q_{\rm tot}$ below
1 or around 1.  The outer north-eastern spiral arm shows $Q_{\rm tot}$
of around 1 and below and few small regions reaching values of about
2.

  In summary, we see three systematic trends in $Q_{\rm tot}$:
  $\bullet$ a slight decrease from the central parts to the outer
  regions. In the spiral arms, peak values decrease from $\sim 1.5-3$
  to $\sim 0.5-1$.  Note that in Paper\,I no systematic change of the
  gas depletion times from the inner to the outer parts was found.
  $\bullet$ Secondly, at a fixed radius, $Q_{\rm tot}$ approaches the
  regime of instability in the arms, while the interarm regions are
  stable. $\bullet$ Thirdly, the outer disk beyond the outer spiral
  arms the disk is stable as well.

  A similar study at higher spatial resolution of the total Toomre
  parameter combining the stellar and gaseous contribution $Q_{\rm
    tot}$ has been done in the LMC by \citet{MacLow2007} using the
  assumption of a constant velocity dispersion. As the LMC has no
  prominent spiral structure, the $Q_{\rm tot}$-map by
  \citet{MacLow2007} also shows no spiral arms with enhanced
  instability. They also stress the importance of the stellar
  contribution for studying the gravitational stability.  Correlating
  the locations of massive young stellar objects (YSOs) in the LMC
  with the gravitational stability, \citet{MacLow2007} find that 85\%
  of the YSOs lie inside gravitationally unstable regions.  This
  underlines the importance of gravitational stability of the disk for
  large-scale star formation in the LMC.

  Both galaxies, the LMC and M51, show large regions near the critical
  value for collapse indicating possible sites for star formation.
  The observation that $Q_{\rm tot}$ is near 1 in large areas of the
  disk is notable and was found previously in radial averages of
  several other galaxies \citep{Boissier2003}. This has led
  several authors, e.g.  \citet{Combes2001}, to speculate that
  self-regulation drives these galaxies near the threshold of
  gravitational stability.

We will briefly sketch the feedback-cycle in the following. The
birthplaces of stars are deeply embedded in dense cores of molecular
clouds formed through gravitational contraction and collapse of
molecular material after loosing much of their turbulent support.
% Once massive star formation has set in, the dispersion in the cloud is
% increased again due to e.g. outflows and the expansion of
% \HII-regions. {\bf CHECK}
%
Both, bipolar outflows and expanding \HII-regions from young stars,
stellar winds and supernovae explosions from evolved stars, will
introduce mechanical energy into the interstellar medium, locally
enhancing the dispersion of the gas, increasing the Q-value. In
addition, ionizing radiation from young stars will lead to
photo-dissociation of the surface regions and the formation of atomic
gas from the molecular material. These feedback mechanisms are
supplemented by the density wave, enhancing cloud-cloud collisions due
to orbit crowding in spiral arms, also leading to an increase of the
velocity dispersion of the molecular gas inside the arms. Star
formation in the remaining molecular gas must await dissipation of the
enhanced macroscopic turbulence to cascade down to smaller scales and
dissipate. Globally, this self-regulation may explain why Q stays near
1.

In the interarm regions the gas stays stable (Q$_{\rm tot}>1$) as no
stars are formed and the feedback mechanisms are not at work.  The
regions critical to collapse in the outer spiral arms coincide with
the finding of increased star formation efficiency in these areas by
\citet{gb1993a} based on H$\alpha$ data. 

\citet{Shetty2008} use hydrodynamic simulations, including feedback to
study star formation and the structure of galactic disks.  They find
that a simple feedback mechanism with one star formation event per
cloud can not sustain a global spiral pattern in galactic disks. This
indicates that the modeling of feedback needs to be improved.
\citet{Quillen2008} explain episodic star formation in galaxies by
strong feedback mechanisms using numerical models.

In recent 3D smoothed particle hydrodynamics simulations of isolated
disk galaxies, \citet{li2005,li2006} confirmed the major role of
gravitational stability in the process of star formation and were able
to examine a threshold surface density. This evidence for the Toomre
formalism in their simulations is also dominating any magnetic effects
on star formation. Magnetic fields slow the collapse of the gas but do
not stop it \citep{Shetty2008}.

\section{Summary} %%%%%%%%%%%%%%%%%%%%%%%%%%%%%%%%%%%%%%%%%%%%% 

Using our complete $^{12}$CO 2--1 map of M51 combined with new HI VLA
data from the THINGS team \citep{Walter2005,Walter2008}, we created a
total gas density map and a ratio map of molecular and atomic surface
density.  The total gas density exhibits an underlying exponential
disk similar to the one found in 850 $\mu$m dust continuum emission by
\citet{Meijerink2005}.  Maps of $^{12}$CO 2--1 velocity field, line
widths and \HI\ line widths were presented. Combining these
information the combined gaseous and stellar Toomre Q-parameter was
discussed radially and locally in the disk of M51. The main results of
this analysis are:

\begin{itemize}  

\item The averaged total gas density on the spiral arms is at around
    30 \msun\,pc$^{-2}$ which is a factor of 3 higher compared to the
    interarm regions. 

  \item The ratio of molecular to atomic gas surface density is
    highest in the inner spiral arms at around 50 decreasing to 0.2 in
    the outer spiral arms. This ratio shows a power-law dependence on
    the hydrostatic pressure $P_{\rm hydro}$ with an index of 0.87,
    similar to that found in other nearby galaxies \citep{Blitz2006}.

  \item We fit an underlying exponential disk to the total gas density
    data with a scale length of 7.6 kpc containing 55\% of the total
    mass. This is comparable to an underlying exponential disk fitted
    to the 850$\mu$m dust continuum map by \citep{Meijerink2005}. The
    gas-dust ratio in the exponential disk stays nearly constant with
    galactocentric radius at around 25. The radially averaged map of
    the gas-to-dust ratio in the residual data shows a variation
    between 80 and 200.

  \item The $^{12}$CO 2--1 velocity field shows distinct differences
    with respect to a purely rotational velocity field. We discuss a
    map of these residual or streaming motions. The observed width
    of the $^{12}$CO 2--1 lines decrease from 100\,km s$^{-1}$ in the
    central part to less than 15\,km s$^{-1}$ in the outer regions. The
    widths of the \HI\ spectra show similar values, but a
    significantly different distribution. While the CO widths show
    higher dispersions in the spiral amrs, the \HI\ shows the maximum
    line widths in the interarm regions and lower values on the spiral
    arms.

  \item We estimate the gravitational stability using a combined
    stellar and gaseous Toomre parameter Q$_{\rm tot}$. The impact of
    the stellar component on the stability is significant reducint the
    Q parameter. It decreases by up to 70\% towards the threshold for
    stability Q$_{\rm total}=1$. Q$_{\rm total}$ varies between 5 in
    the interarm regions decreasing to values below 1 on the spiral
    arms.  In general, the spiral arm regions are closer to the
    threshold for gravitational instability.  The obtained values
    close to the threshold of 1 indicate self-regulation in the disk
    of M51. The self-regulation mechanism was discussed on theoretical
    grounds by \citep[e.g.][]{Combes2001,Shetty2008}. Similar studies
    based on extragalactic observations in M83 \citep{lundgren2004_II}
    or the LMC \citep{MacLow2007} obtained also Q-values close to the
    threshold for stability.

\end{itemize}  

%%%%%%%%%%%%%%%%%%%%%%%%%%%%%%%%%%%%%%%%%%%%%%%%%%%%%%%%%%%%%%%%%%%%%%%%%%%%%% 
\begin{acknowledgements}
  We thank F.\,Walter for providing us the THINGS VLA \HI\ data and we
  thank R.\,Meijerink, R.\,Tilanus, and F.\,Israel for providing us
  the SCUBA 850\,$\mu$m data of M51.  MH acknowledges support from the
  Bonn-Cologne Graduate School of Physics and Astronomy (BCGS).  This
  work is also financially supported in part by the grant SFB\,494 of
  the Deutsche Forschungsgemeinschaft, the Ministerium f\"ur
  Innovation, Wissenschaft, Forschung und Technologie des Landes
  Nordrhein-Westfalen and through special grants of the Universit\"at
  zu K\"oln and Universit\"at Bonn.
\end{acknowledgements} 

\bibliographystyle{aa} %%%%%%%%%%%%%%%%%%%%%%%%%%%%%%%%%%%%%%%%%%%%%%%%%% 
\bibliography{aamnem99,p_toomre_bib} % p_galaxies_bib.bib 

\end{document}